\documentclass[aps,showpacs,twocolumn,floats,superscriptaddress]{revtex4}
\usepackage{graphicx}
\usepackage{amsmath}
\usepackage{color}

\newcommand{\be}{\begin{equation}}
\newcommand{\ee}{  \end{equation}}
\newcommand{\ba}{\begin{eqnarray}}
\newcommand{\ea}{  \end{eqnarray}}

\begin{document}
\title{Experimental implementation of assisted quantum adiabatic passage in a single spin}

\author{
Jingfu Zhang$^1$, Jeong Hyun Shim$^1$, Ingo Niemeyer$^1$, T. Taniguchi$^2$,
T. Teraji$^2$, H. Abe$^3$, S. Onoda$^3$, T. Yamamoto$^3$, T. Ohshima$^3$, J. Isoya$^4$  and Dieter Suter$^1$\\
 $^1$Fakult\"{a}t Physik, Technische Universit\"{a}t Dortmund,
 D-44221 Dortmund, Germany\\
$^2$National Institute for Materials Science,
1-1 Namiki, Tsukuba, Ibaraki, 305-0044 Japan \\
$^3$Japan Atomic Energy Agency, 1233 Watanuki, Takasaki, Gunma,
370-1292 Japan\\
$^4$Research Center for Knowledge Communities, University of
Tsukuba, Tsukuba, 305-8550 Japan}

\date{\today}

\begin{abstract}
Quantum adiabatic passages can be greatly accelerated  by a
suitable control field, called a counter-diabatic field, which
varies during the scan through resonance. Here, we implement this
technique on the electron spin of a single nitrogen-vacancy center
in diamond. We demonstrate two versions of this scheme. The first
follows  closely the procedure originally proposed by Demirplak
and  Rice (J. Phys. Chem. A 107, 9937 (2003)).  In the
second scheme, we use a control field whose amplitude is
constant, but its phase varies with time. This version, which we
call the rapid-scan approach, allows an even faster passage
through resonance and therefore makes it applicable also for
systems with shorter decoherence times.
\end{abstract}
\pacs{03.67, 33.35, 76.70} \maketitle

{\it Introduction.--} Controlling quantum systems with high
fidelity is an essential prerequisite in various fields, such as
coherent control of atomic and molecular systems \cite{RMP79.53}
and quantum information processing
\cite{NielsenChuang,StolzeSuter}. The strategies that have been
developed for this purpose include the adiabatic passage
technique, which leads the quantum system along a specific pathway
in such a way that the system always remains in its ground state.
One of the attractive properties of this technique is that the
resulting evolution is robust with respect to some experimental
imperfections \cite{PhysRevA.65.012322}.
Adiabatic passage also is the central part
of the adiabatic model of quantum computation
\cite{Farhi00,Science292.472}, which has been shown to be
equivalent to the more common network model.  In all these cases,
it is essential that the scan duration of the adiabatic passage is
short and the fidelity as high as possible.

The quantum adiabatic theorem guarantees that the system remains
approximately in its ground state if the evolution is sufficiently
slow
\cite{PhysRevLett.101.060403,PhysRevLett.95.110407,PhysRevLett.93.160408}.
However, for all practical applications, the optimal
implementation is reached when the scan time remains short, e.g.
compared to the decoherence time.  A variety of techniques have
been developed, such as exploiting nonlinear level-crossing models
\cite{nonlinear}, amplitude-modulated and composite pulses
\cite{ampmod,composite11}.

 In a recent development
\cite{J.Phys.Chem.A107.9937,Berry09,chenPRL10}, it was shown that
the system can remain exactly in its ground state, without
undergoing transitions, if an additional control field, the
so-called the counter-diabatic (CD) field is introduced. This
strategy was recently implemented in an atomic Bose-Einstein
condensate \cite{naturephys.8.147}.

In multilevel systems, stimulated Raman transitions can be driven in such a way
that populations are transferred adiabatically between two states
\cite{RMPShore98,VitanovRew01} .
Adiabatic transfers have also been extended to nonlinear systems,
where the theoretical analysis become significantly more complicated
\cite{nonlinear1,nonlinear2}.

In this Letter, we report another experimental implementation of
the assisted adiabatic passage (AAP), using a single nitrogen
vacancy (NV)-center in diamond
\cite{rewievNVC78,NJP11,PhysRevB.85.205203}.  The NV- centers are
point defects, each of which consists of a substituational
nitrogen
 adjacent to a vacancy.
 The NV-centers can exhibit
attractive quantum properties even at room temperature
\cite{Gaebel06,P.Neumann10,Nature.484.82}.
The potential applications of the NV-centers  include quantum
metrology \cite{Qmetrology2,Qmetrology3,nnano12,Qmetrology4} and
quantum computing
\cite{QCNVCrew1,JPhys18.S807,nature.464.45,Naturemat.9.468}.
Various techniques for implementing high-fidelity coherent control
of the NV-centers have been developed recently
\cite{singleDec1,Nature.484.82,PhysRevLett.109.070502,nnano12}. In
our experiment, we implement the AAP on an electron spin
transition of the NV-center, using resonant microwave fields as
controls.  In contrast to the previous implementation
\cite{naturephys.8.147}, where the coherent control was applied to
an ensemble of atoms,
 our experiment is implemented on a single
spin, with a potential advantage in encoding quantum information
into qubits in building quantum computers. The fidelity of the
passage is sufficiently high that we can perform multiple rounds
of the passage in opposite directions. The results show good
agreement between theory and experiment.

\begin{figure}[bt]
\includegraphics[width=3in]{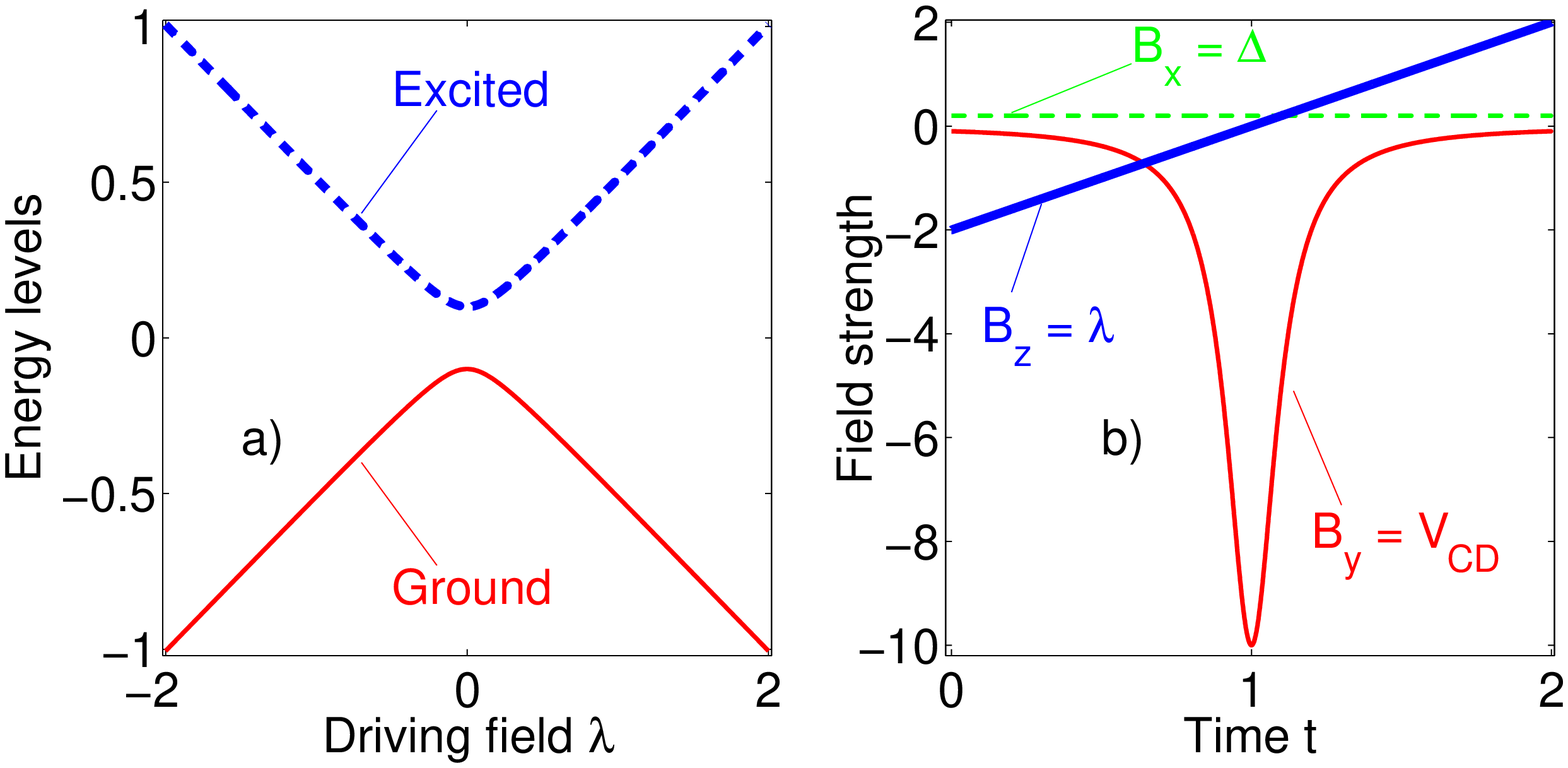} 
\caption{(color online). Characteristics of the single qubit model
for $\Delta=0.2$ and  $b = 2$. (a) shows the energy levels and (b)
the time dependence of the three field components.}
\label{figmodel}
\end{figure}

{\it Model.--} In close analogy to the Landau-Zener model
\cite{LZ1,LZ2,PhysicsReports.492.1}, we describe the AAP with the
Hamiltonian
\begin{equation}\label{ham2}
    H_{LZ}(t)=\lambda(t) I_{z} + \Delta I_{x} ,
\end{equation}
where the $I_{x,z}$ are spin operators, and $\lambda(t)$ and
$\Delta$ are dimensionless fields applied along the $z$ and $x$
directions. This Hamiltonian is a model for an arbitrary two level
system and plays a prominent role in various fields of physics,
such as in coherent control
\cite{naturephys.8.147,PhysRevLett.109.060401},  quantum
criticality
\cite{PhysRevLett.95.035701,PhysRevLett.105.240406,Zurek12},
 and even in many-body systems\cite{Garanin}.
The energy eigenvalues are
$\pm\sqrt{\lambda^{2}(t)+\Delta^{2}}/2$, and the instantaneous
ground state is
\begin{equation}\label{insgrouns}
   |g(t)\rangle = \sin [\theta(t)/2]|0\rangle - \cos[\theta(t)/2] |1\rangle,
\end{equation}
 where $\tan \theta(t) = \Delta/\lambda$,
$\theta\in [0,\pi]$, and $|0\rangle$ and $|1\rangle$ denote the
eigenstates of $I_{z}$ with eigenvalues $\pm1/2$, respectively.
The minimal gap between the
two levels is $\Delta$. 
 Figure
\ref{figmodel} a) illustrates the energy levels for $\Delta=0.2$.

The original assisted adiabatic passage model
\cite{J.Phys.Chem.A107.9937,Zurek12} starts from a scan from
$-\infty$ to  $+\infty$. For the experimental implementation, we
have to restrict ourselves to a finite range, which we write as
$[-b, b]$. If the control field is scanned linearly in time from
$-b$ to $b$,
\begin{equation}\label{Bzt}
    \lambda(t)=b(t - 1),
\end{equation}
where 
$t\in [0,2]$.
The scheme can then be implemented for any non-zero
value of $\Delta$ by introducing a CD field that is perpendicular
to both the $x$-  and  $z$-components  of the field:
\begin{equation}\label{HamCD}
    H_{CD}(t) =  V_{CD}(t) I_y
\end{equation}
where
\begin{equation}\label{cddriving}
    V_{CD}(t) =
    -(d \lambda/dt) \Delta/[\Delta^2+\lambda(t)^2],
\end{equation}
where  the  rate of change of $\lambda(t)$ is $d\lambda/dt=b$ for
the field of Eq. (\ref{Bzt}). The total field for the AAP is thus
a vector $\vec{B}(t) = [\Delta$, $V_{CD}(t)$, $\lambda(t)]$.
Figure \ref{figmodel} b) shows the time dependence of the three
components for $b = 2$.


{\it Experimental protocols and results.--}  For the experimental
test, we used the electron spin of a single NV-center in $^{12}$C
enriched diamond.
 The reduced number of
$^{13}$C nuclear spins results in long relaxation times, with
$T_2^\ast > 100 \, \mathrm{\mu s}$.

The hyperfine coupling between the electron and the $^{14}$N
 nuclear spin is  $\approx 2.1$ MHz, see the spectrum shown in
 Figure 1 a) in the Supplementary Material \cite{SM}.
 For the present experiments, we use the subspace of this system that is spanned by the states
 $m_S = 0, +1$ of the electron spin, $m_I = 0$ of the nuclear spin.
 This subsystem can be excited with excellent selectivity
 if the amplitude of the microwave field remains well below the hyperfine coupling constant.
 We therefore will not consider the nuclear spin state in the
 following,  see the spectrum shown in
 Figure 1 b) in the Supplementary Material \cite{SM}.

For the AAP, the system should be initialized into the ground
state  $|g(0)\rangle$ of $H_{LZ}(0)$. In the experiment, we
initialize the system by a laser pulse into the state $|0\rangle$,
whose overlap with $|g(0)\rangle$ is $\sin[\theta(0)/2]$. In the
experiment, we use $-3 \geq \lambda(0)/\Delta \geq -20 $, which
results in overlaps of $[0.9871, 0.9997]$. For our purpose,  this
is sufficiently close to unity.

After initialization, the system evolves under the time-dependent Hamiltonian
\begin{equation}\label{Hamwhole}
 H(t) = H_{LZ}(t) + H_{CD}(t)
\end{equation}
into the state
\begin{equation}\label{stateU}
  |\psi(t)\rangle = \widetilde{U}(t) |g(0)\rangle,
\end{equation}
where  $\widetilde{U}(t)$ represents the propagator generated by
$H(t)$. During and after the scan, we read out the state of the
system by a second laser pulse, which again projects the system
onto the state $|0\rangle$. We write the probability of finding
the system in this state as $P_{|0\rangle} =
|\langle0|\psi(t)\rangle|^2$. The detailed description of the
system and the experiment setup is given in the Supplementary
Material \cite{SM}.

In the actual experiment, the  fields $\Delta$, $V_{CD}$ and $\lambda$ act on the spin in a rotating reference frame.
Writing
$$
U_r = e^{-i\xi(t)I_{z}}
$$
for the transformation from the laboratory-frame to the rotating
frame, the Hamiltonians of the two frames are related as
\begin{eqnarray}
H^\text{rot} &=& U_r H^\text{lab}  U_r^\dag + i \dot U_r U_r^\dag \nonumber \\
H^\text{lab} &=& U_r^\dag  H^\text{rot}   U_r - i  U_r^\dag \dot U_r  .
\label{e.trafo}
\end{eqnarray}
The laboratory-frame Hamiltonian thus has the field components
\begin{eqnarray*}
\omega_x(t) &=& \Delta \cos\xi(t) + V_{CD}(t) \sin \xi(t) \\
\omega_y(t) &=&  -\Delta \sin\xi(t) + V_{CD}(t) \cos\xi(t) \\
 \omega_z &=& \lambda(t) -  [d \xi(t)/dt].
\end{eqnarray*}
This Hamiltonian must match the experimentally available Hamiltonian, whose general form is
$$
H^\text{exp} = - \omega_0 I_z + 2 \omega_1(t) I_x.
$$
Accordingly, we must have
$$
\xi(t) =   \omega_0 t +  \int_0^t \lambda(t') dt' ,
$$
which defines our rotating frame transformation.
For the transverse field components, we invoke the rotating field approximation, which allows us
to set
$$
\omega_1(t) = \Delta \cos\xi(t) + V_{CD}(t) \sin \xi(t)
$$
and ignore the $y$-component.
The amplitude of the field is therefore
$$
|\omega_1(t)| = \sqrt{\Delta^2 + V_{CD}(t)^2}.
$$
Experimental limitations define a maximum possible field
amplitude, which we designate as $\Omega$. In the present
experiment, it is determined by the requirement that no
transitions of other nuclear spin states are excited, and we found
a value of $\Omega \approx 2\pi \cdot 0.2 \, \mathrm{MHz}$ to be a
suitable compromise. The maximum field amplitude is reached at
\begin{equation}\label{scalingana}
 \Omega = \sqrt{\Delta^2 + [b^2/\Delta^2]}/s_a,
\end{equation}
where we have defined the scale factor $s_a$, which  converts the
dimensionless quantities $\Delta$ and $b$ into actual field
amplitudes (in Hz) and defines the scan duration
\begin{equation}\label{scandurationana}
  \tau_s = 2 s_a.
\end{equation}

\begin{figure}[bt]
\includegraphics[width=3in]{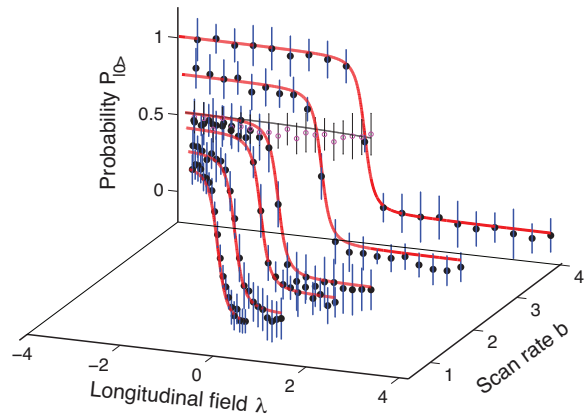} 
\caption{(color online). Experimental results obtained with the
analog implementation of the AAP. The individual experiments
correspond to  $b = 0.6$, $1$, $1.6$, $2$, $3$ and $4$. The
experimental data are shown as filled circles, and the
corresponding error bars were obtained by repeating the experiment
10 times. The curves show the theoretical result for an ideal
scan. The empty circles with the almost horizontal line show the
result for a reference experiment without CD field. The 2
dimensional representation is shown in the Supplementary Material
\cite{SM}. }\label{FigRob}
\end{figure}

 Figure \ref{FigRob} shows the experimental results of the AAP for the scan rates
 $b = 0.6$, $1$, $1.6$, $2$, $3$ and $4$.
 For these parameters, we can approximate $s_a \approx b/( \Omega \Delta)$.
 For $b=0.6$, the scan duration becomes  $4.77 \, \mathrm{\mu s}$
 and for $b=4$ it is $31.83 \, \mathrm{\mu s}$.
 The $z$-component is always scanned from $-40$ to $+40$ $\mathrm{kHz}$
 (in frequency units), while the $y$-component
 (the CD field) reaches a maximal amplitude of $200$ $\mathrm{kHz}$ at the anticrossing point.
 The $x$-component of the field in the rotating frame is $13.3$ $\mathrm{kHz}$ for $b=0.6$
 and $2.0$ $\mathrm{kHz}$  for $b=4$.
Filled circles show the experimentally measured overlaps of the
state with the ground state $|0\rangle$. The error bars (1
standard deviation) were determined by repeating each experiment
10 times. The solid lines, which agree very well with the
experimental data points, represent the theoretical behavior. For
comparison, we also show one data set that was obtained without CD
field, for a scan rate of $b=2$. These data points are represented
by the empty circles and the corresponding theoretical curve can
be  approximated as  a horizontal line going through the
experimental points. Clearly, a passage without CD field results
in an  almost  completely diabatic transfer.

Heisenberg's uncertainty relation limits the speed of every (adiabatic or not) state-to-state transfer
for a given field strength. It is thus possible to increase the speed by using higher field strengths.
On the other hand, the maximum field strength is limited by the properties of the system as well
as by experimental limitations.
Within these experimental limitations, we now look for a scheme that remains close to the
original proposal, but minimizes the overall duration of the scan without exceeding
a field strength that is dictated by the experimental conditions.
In the following, we will refer to this approach as the rapid-scan.

We determine the required control fields by dividing the overall
evolution $\widetilde{U}(t)$ into $N$ segments
\cite{PhysRevA.65.042323,PhysRevA.71.012307}, each of duration
$\delta$, with the total duration of the sequence $N \delta = 2$.
The evolution within each segment is $U_m = \mathcal{T}\exp[- i
\int_{(m-1)\delta}^{m \delta} H(t) dt] \approx \exp[- i \delta H(m
\delta)]$, where $\mathcal{T}$ denotes the Dyson time ordering
operator. Each segment $U_m$ was implemented as a rectangular
pulse with Hamiltonian  $H_m = H(m\delta) / s_m$, whose field
amplitude had the constant value $\Omega$. $s_m$ is the scaling
factor for the segment. The duration of the segment can therefore
be reduced by this factor, compared to $\delta$, to $\tau_m = s_m
\delta$. Clearly, the reduction of the duration is only limited by
the  available field strength.


Using the transformation (\ref{e.trafo}), we can calculate the required laboratory-frame
Hamiltonian
$$
H_m^{{\rm lab}}= - \omega_0 I_z + 2 \Omega I_x \cos(\omega_m t +
\phi_m)
$$
and the required duration $\tau_m$, where
  $ t \in [\sum_{j=1}^{m-1} \tau_{j}, \sum_{j=1}^{m} \tau_{j}]$.
   The scaling factor
\begin{equation}\label{scalefactordig}
s_m = \tau_m/\delta =  \sqrt{\Delta^2 + V_{CD}^2(m\delta)}/\Omega
\end{equation}
is now different for every segment.
The angular frequency $\omega_m$ and the phase $\phi_m$ become
$$
\omega_m = \omega_0 + [\lambda(m\delta)/s_m],
 \hspace{0.5cm}
 \tan\phi_{m} = - V_{CD}(m\delta)/\Delta.
$$
In the Supplementary Material \cite{SM}, we show the explicit values of these parameters
for each step.

\begin{figure}[bt]
\includegraphics[width=3in]{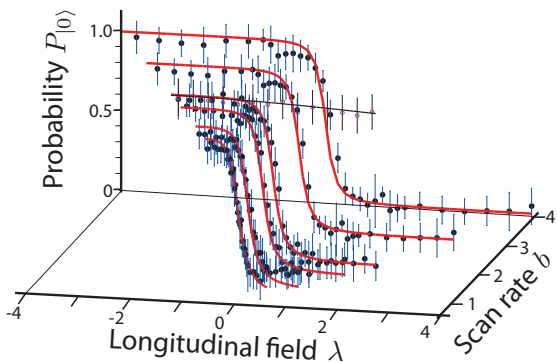} 
\caption{(color online). Experimental results obtained with the
rapid-scan approach. For details see the caption of Figure
\ref{FigRob}.
The 2 dimensional representation is shown in the
Supplementary Material \cite{SM}.  }\label{FigRob1}
\end{figure}

Figure \ref{FigRob1} shows the experimental results.
Here, we used the same  nominal scan rates $b$ as in the analog case,
but split the scan into $N = 56$ segments.
The experimental data points are represented by filled circles, the error bars
were obtained by repeating the experiments 8 times.
Clearly, the experimental data agree very well with the theoretical expectation shown
as the red curves.
The empty circles again represent the reference experiment obtained by setting
$V_{CD} = 0$.

\begin{figure}[bt]
\includegraphics[width=3.5in]{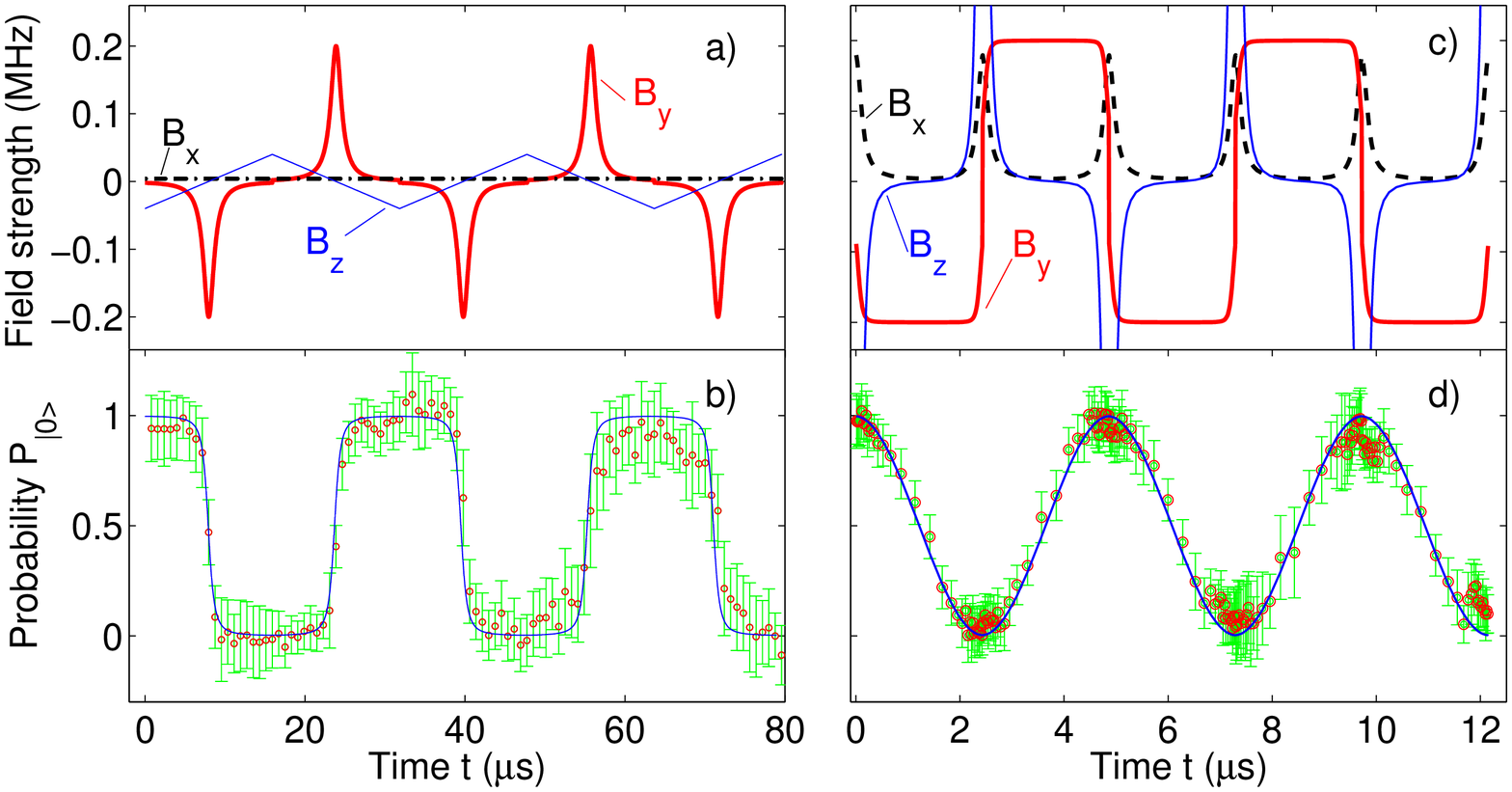} 
\caption{(color online). Multiple assisted adiabatic passages.
 The results obtained in the analog and rapid-scan
approaches are shown in the left and right columns, respectively.
Figures a) and c) show the three components of the applied fields,
while b) and d) show the experimental data, together with the
theoretical curves for the ideal case (zero loss).}
\label{FigmultiRND}
\end{figure}

Given the high fidelity obtained in a single passage through
resonance, we can cycle the system forth and back multiple times.
The reverse passage is obtained by  changing $b$ to $-b$, which
changes the direction of the scan as well as the sign of the CD
field.  The results obtained in the analog and rapid-scan
approaches are shown as the left and right columns in Figure
\ref{FigmultiRND}. Figures a) and c) show the three field
components in the rotating frame, and b) and d) represent the
experimental results for these repetitive scans, as well as the
theoretical curves corresponding to the ideal case. We find very
little loss of population after five passages through resonance.

{\it Discussion.--}
The experimental implementation of  the time-dependent Hamiltonian $H(t)$
always occurs with finite precision, which results in a loss of fidelity.
Experimental contributions to this loss include the precision with which
the shaped pulses are implemented - both in terms of the amplitude
as well as in terms of the time resolution.
In the rapid-scan approach, the number of segments used is an important parameter.
We used numerical simulations of the experiment to estimate these
losses. The results indicate that for the analog scan, finite time
resolution of  $0.25$ ns reduces  the  fidelity by a fraction
of the order of $10^{-8}$. 
In the rapid-scan approach, for $N = 56$
segments, the maximal loss during the AAP occurs near the critical
point at $\lambda = 0$. For the lowest scan rate, with $b = 0.6$,
this loss is of the order of $<10^{-3}$, for the faster scan rate,
$b = 4.0$, it rises to $2.8 \cdot 10^{-2}$. However, these are
mostly intermediate losses, which are recovered during the second
part of the scan: The calculated loss of fidelity at the end of
the
evolution period is $<10^{-4}$. 


For the parameters chosen here, the duration of a single scan varies between
$4.77$ and $31.83$ ${\mathrm \mu s}$ in the analog version and from
$2.0$ to $2.5$ ${\mathrm \mu s}$ in the rapid scan implementations.
They are thus all short compared to the coherence time of
our sample ($T_2 \approx 500 \mu \mathrm{s}$). 

To estimate the speed-up provided by the CD field, we used
numerical simulations of an  unassisted scan, with the same
parameters as the experimental scan in Figure \ref{FigRob}, but
slower scan rates. To reach a fidelity of  0.99, the scan duration
had to be extended to $2.33$ ${\mathrm{ms}}$. This implies that
the assisted scan allows a speedup of more than two orders of
magnitude ($\sim 150$) if a linear frequency scan is used and of
about three orders of magnitude ($\sim 960$) in the rapid passage
(constant amplitude) version.

 The analog and the rapid-scan approaches implement both
the same propagator, but they use a different scaling of the
time-axis.  This allows one to scan very rapidly when the
quantization axis does not change appreciably with the offset.
Most of the speed advantage of the rapid-scan approach is
therefore gained in the region of large detunings (see Figure 3 in
the Supplementary Material). Both experiments relied on a
segmentation of the actual fields for implementation in an
arbitrary waveform generator. The precision with which the scans
can be implemented depend therefore on the amplitude- and
time-resolution of the instrument. In our setup, the minimal
possible time-resolution is 0.25 ns, which is significantly
shorter than the time step used here ($ > 6.4$ ns). According to
numerical simulations, using the full time resolution would reduce
the loss of fidelity due to the segmentation to $<10^{-7}$. The
rapid-scan approach presented here is a first attempt at reducing
the duration of an AAP scan. We are confident that further
improvements are possible, e.g. by using the tools of optimal
control theory.

{\it Conclusion.--}
We have implemented the assisted adiabatic
passages through analog and rapid-scan approaches in a two level
quantum system by controlling a single spin in a NV-center in
diamond. This approach allows a significant increase in the scan
rate compared to the unassisted passage and therefore reduces the
requirements on the decoherence time of the system to which it is
applied.
Like in the unassisted case, the scan has to be slower if the minimum gap is small.
If the scan is
performed linearly in time, the total duration also increases with
the scan range. However, with the rapid-scan approach that we
introduced here, the scan range can be increased arbitrarily with
very little time-penalty. Our experiment results illustrate the
excellent coherent control that can be achieved for the spins of
NV-centers. These results should be helpful for all applications
requiring quantum adiabatic passages, such as implementing
geometric gates for quantum computation \cite{Geom12}, adiabatic
control in interacting two level systems
\cite{PhysRevLett.109.043002} or adiabatic quantum computing
\cite{Science292.472}.

This work is supported by the Heinrich Hertz Foundation, and the
DFG through grant Su 192/27-1.


\end{document}